\documentclass[sn-nature,iicol]{sn-jnl}

\usepackage{graphicx}%
\usepackage{multirow}%
\usepackage{amsmath,amssymb,amsfonts}%
\usepackage{amsthm}%
\usepackage{mathrsfs}%
\usepackage[title]{appendix}%
\usepackage{xcolor}%
\usepackage{textcomp}%
\usepackage{manyfoot}%
\usepackage{booktabs}%
\usepackage{algorithm}%
\usepackage{algorithmicx}%
\usepackage{algpseudocode}%
\usepackage{listings}%

\newcommand{\add}[1]{{\color{black}#1}}

\begin{document}

\title[Article Title]{\add{Picosecond-scale} Heterogeneous Melting of Metals \add{at} Extreme Non-equilibrium States}

\author[1,2,3]{Qiyu Zeng}
\author*[1,2,3]{Xiaoxiang Yu}\email{xxyu@nudt.edu.cn}
\author[1,2,3]{Bo Chen}
\author[1,2,3]{Shen Zhang}
\author[1,2,3]{Kaiguo Chen}
\author*[1,2,3]{Dongdong Kang} \email{ddkang@nudt.edu.cn}
\author*[1,2,3]{Jiayu Dai}\email{jydai@nudt.edu.cn}

	\affil[1]{College of Science, National University of Defense Technology, 410073 Changsha, Hunan, China}
	\affil[2]{Hunan Key Laboratory of Extreme Matter and Applications, National University of Defense Technology, 410073 Changsha, Hunan, China}
    \affil[3]{Hunan Research Center of the Basic Discipline for Physical States, National University of Defense Technology, Changsha 410073, China}


\abstract{
Extreme electron-ion non-equilibrium states, generated by ultrafast laser excitation, lead to melting processes that are fundamentally different from those under conventional thermal equilibrium and remain not fully understood. Through neural network-enhanced multiscale simulations of tungsten and gold nanofilms, we identify electronic pressure relaxation as critical \add{to} heterogeneous phase transformations. \add{This nonthermal} expansion generates \add{a} density decrease that enable surface-initiated melting far below equilibrium melting temperatures, creating electronic pressure-driven solid-liquid interface propagation at a high speed of $2500 \ \rm{m\ s^{-1}}$ — tenfold faster than that of thermal heterogeneous melting mechanisms. Simulated time-resolved X-ray diffraction signatures distinguish this nonthermal expansion from thermal expansion dynamics driven by thermoelastic stress. These results establish hot-electron-mediated lattice destabilization as a universal pathway for laser-induced structural transformations, providing new insights for interpreting time-resolved experiments and controlling laser-matter interactions.
}




\maketitle

\section{Introduction}\label{sec1}

Ultrafast laser excitation has emerged as a transformative tool for probing and manipulating matter under extreme nonequilibrium conditions, enabling breakthroughs from attosecond spectroscopy to high-precision nanofabrication \cite{guan2022,  gattass2008}. A hallmark of these interactions is the transient electronic excitation that establishes a pronounced temperature disparity between electrons and lattices ($T_{\rm e} \gg T_{\rm i}$) — a nonequilibrium regime where hot electrons coexist with a cold ionic framework. This state, which differs from thermal equilibrium ($T_{\rm e} = T_{\rm i}$), dynamically reconfigures interatomic interactions: charge density redistribution modulates potential energy surfaces (PES), altering bonding forces and energy barriers \cite{fritz2007, liu2022}. Such processes deviate from equilibrium thermodynamic pathways, raising fundamental questions about how materials evolve when energy deposition outpaces thermal relaxation, particularly in laser-induced melting.

The isochoric hypothesis, which assumes negligible volume changes during sub-picosecond heating events, has been the cornerstone of our understanding of material response upon laser excitation. Within this picture, a plethora of femtosecond phenomena have been uncovered: bond hardening \cite{recoules2006, smirnov2020}, phonon softening \cite{arnaud2013} and dynamic lattice instabilities \cite{zijlstra2008, lian2016}. These processes occur on timescales of $\sim 10^2$ fs, where inertial stress confinement ensures a quasi-constant density. However, emerging evidence reveals a significant shift in understanding material response at picosecond timescales: hot electrons can exert an additional pressure component $p_{\rm e}$—besides the thermoelastic stress—through thermal kinetic energy and quantum degeneracy of thermalized electrons \cite{khakshouri2008, bevillon2014, zeng2023}, resulting in a volume change that may challenge the isochoric assumption. In gold, for instance, isobaric calculations predict phonon softening and nonthermal melting \cite{daraszewicz2013, medvedev2020}, in stark contrast with the isochoric hardening evidenced in sub-ps X-ray diffraction experiments \cite{descamps2024}. This divergence highlights a critical knowledge gap: How does the accumulation of electronic pressure (during $\sim 10^2$ fs) and its relaxation (on $\sim 10^1$ ps timescale) affect structural transformations?

Addressing this issue is challenging because the laser-driven process involves several coupled physical processes across different scales, including laser excitation that modifies the PES, non-adiabatic electron-ion energy exchange, and atomic-scale structural responses \cite{lian2016, munoz2023}.
Experimentally, the above issues, along with inherent limitations in temporal and spatial resolution, confounds the interpretation of measured data \cite{ernstorfer2009, mo2018heterogeneous, molina2022, arefev2022}. 
Theoretically, the laser-driven process presents dual requirements that existing methodologies struggle to reconcile: \textit{ab initio} accuracy to capture the hot-electron-modified PES governing $p_{\rm e}$, and atomic-scale resolution across experimentally relevant thickness (tens of nanometers) to track stress wave propagation and phase transition. 
\textit{Ab initio} accuracy and large-scale simulation size significantly constrains the efficacy of existing methodologies, such as classical molecular dynamics combined with two-temperature model \cite{ivanov2003combined, zeng2020structural} or real-time time dependent density functional theory \cite{lu2021, xu2022, li2024cage}, thereby hindering a thorough exploration into the real-time response of laser-excited material without any prior constraints.   

Here, we unify \textit{ab initio} accuracy with large-scale molecular dynamics using recently-developed hybrid atomistic-continuum approach, known as two-temperature model coupled deep potential molecular dynamics (TTM-DPMD) \cite{zeng2023}. By simulating free-standing tungsten (W) and gold (Au) nanofilms, we uncover the crucial role of electronic pressures in triggering heterogeneous structural transformations at picosecond scales. Time-resolved X-ray diffraction signatures further distinguish this mechanism from thermal expansion, offering experimental validation pathways. 

\section{Results}\label{sec2}

\textbf{\textit{Ab initio} modeling of laser-driven dynamics. }
Within the TTM-DPMD framework, the electron temperature dependent deep neural network (ETD-DNN) is implemented to capture the \textit{ab initio} laser-excited PES while maintaining computational efficiency. With ETD-NN, the nonthermal contribution to total energy, forces, and pressure at elevated electronic temperature $T_{\rm e}$ is inherently incorporated \cite{supple}. 
To ensure the accuracy of our neural network model, we have validated the thermophysical and vibrational properties of W and Au under both equilibrium ($T_{\rm e}=T_{\rm i}$) and non-equilibrium conditions ($T_{\rm e}\ne T_{\rm i}$) \cite{zeng2023,supple}. Especially, we reproduced the phonon softening in W along the $\rm{H-N}$ and $\rm{N-\Gamma}$ paths in the first Brillouin zone at elevated electron temperatures up to $20,000\ \rm{K}$, consistent with previous predictions \cite{giret2014} (see \add{Suppl. Fig.1}). Under more severe non-equilibrium state ($T_{\rm e} = 22,000\ \rm{K}$), an imaginary phonon mode at the $\rm{N}$-point was observed, indicating a possible solid-solid phase transition. Here we mainly focus on the moderate non-equilibrium condition below the $T_e=20,000\ \rm{K}$ to exclude the influence of lattice-instability-driven nonthermal phase transition.

\begin{figure*}[htbp]
\centering
\includegraphics[width=1.0\linewidth]{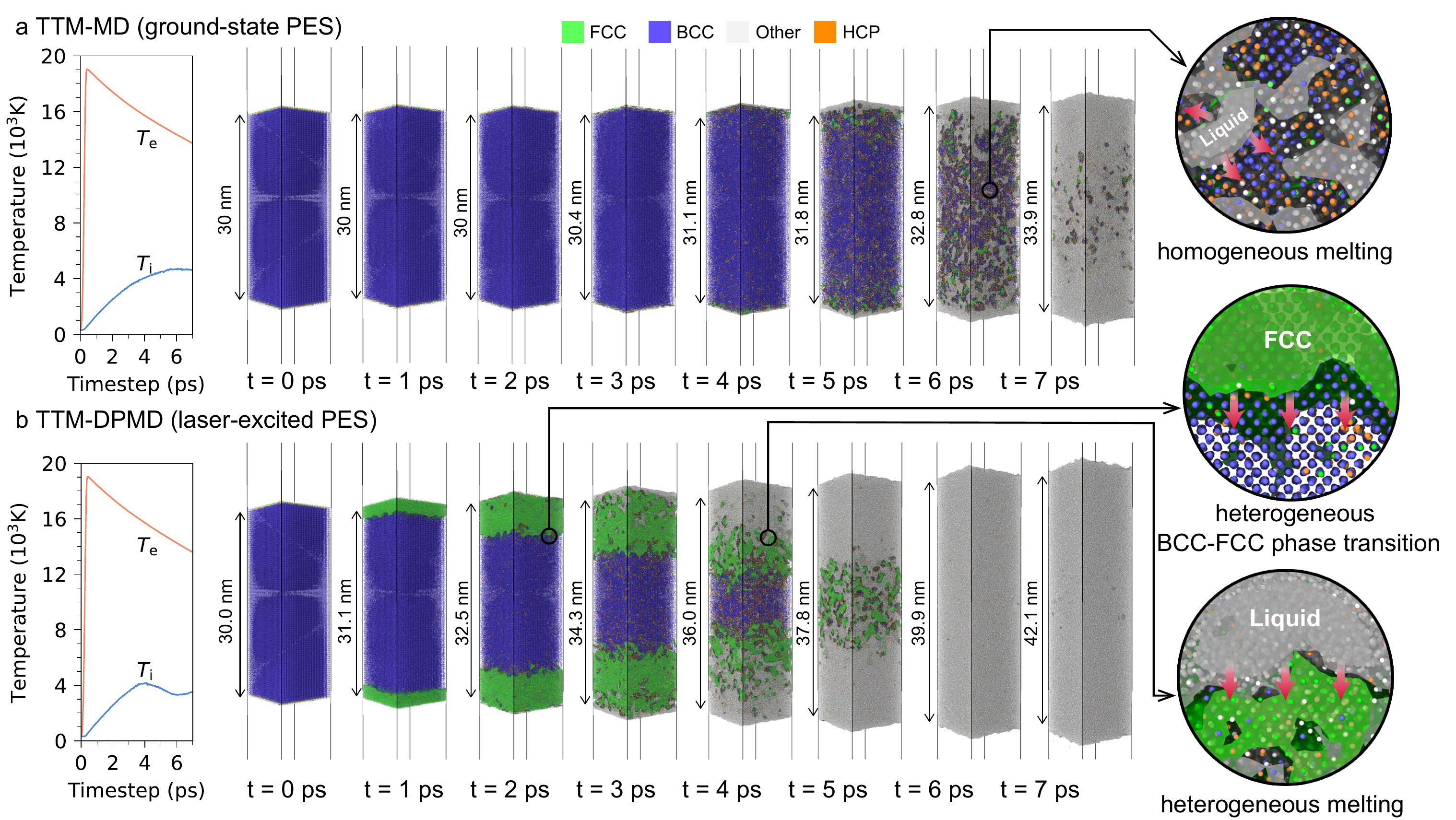}
\caption{\textbf{Diverging melting dynamics in laser-excited tungsten.} Under absorbed laser fluence of $120\ \rm{mJ\ cm^{-2}}$,  (a) Conventional TTM-MD prediction with ground-state PES $A(\mathcal R)$ showing homogeneous melting \add{governed} by electron-phonon coupling. (b) Our TTM-DPMD results with laser-excited PES $A
(\mathcal R, T_{\rm e})$ revealing electronic pressure-driven heterogeneous melting. The local structures are identified by the polyhedral template 
matching (PTM) method \cite{larsen2016}, and polyhedral surface meshes around FCC-type (green) 
and amorphous-type (gray) particles are constructed to highlight the heterogeneity in lattice symmetry. Red arrows indicate the propagation direction of phase transition (solid-liquid or BCC-FCC) interfaces. \add{The atomic configurations are visualized by OVITO software \cite{ovito}.}} 
\label{fig:1}
\end{figure*}

The efficiency of TTM-DPMD approaches enables a full-scale \textit{ab initio} description where the geometry of the sample is compatible with the experimental conditions. Here, we choose 30-nm-thick W nanofilm \cite{mo2019} and 35-nm-thick Au nanofilm \cite{mo2018heterogeneous} as the target sample. 
Since the foil thickness is comparable to the mean free path of excited electrons ($\sim 33\ \rm{nm}$ in W and $\sim 100\ \rm{nm}$ in Au), 
the ballistic electron transport and reflux inside the foil produces uniform laser energy deposition \cite{chen2012flux}.
Free boundary condition is applied to the laser incident direction ($z$-axis) to allow free surface response to the relaxation of internal stress (including both electronic and thermoelastic contributions). Periodic boundary conditions are applied in the lateral directions to simulate the experimental conditions where laser spot diameter is large (hundreds of microns) compared to the depth of laser energy deposition \cite{chen2018interatomic}. 
More details about TTM-DPMD simulation are provided in the Method Section.

\textbf{Ultrafast heterogeneous melting dynamics.} 
Here we present structural transformation of laser-excited W subjected to an absorbed laser fluence of $120\ \rm{mJ\ cm^{-2}}$ (duration of laser pulse set to 130 fs). This fluence, although insufficient to trigger dynamical instability in the BCC lattice, reveals significant alterations in melting behavior due to the presence of hot electrons.

For comparison, we firstly discuss the purely thermal process obtained from a traditional TTM-MD method with ground-state PES that does not incorporate laser-modification (Fig.\ref{fig:1}a). After laser energy deposition, the system reaches maximum electron temperature $T_{\rm e}=19,050\ \rm{K}$ while the lattice remains cold ($T_{\rm i}=300\ \rm{K}$). The lattice temperature quickly increases due to electron-phonon energy exchange. Once lattice temperature exceeds the limit of lattice thermal stability, the nucleation of liquid region inside the foil is triggered, known as homogeneous melting mechanism \cite{ivanov2003combined}. Within the first 5 ps, surface expansion remains limited to moderate volumetric change (\add{Suppl. Fig.2a}), as constrained by the timescale of thermal pressure buildup.

\begin{figure}[htbp]
\centering
\includegraphics[width=1.0\linewidth]{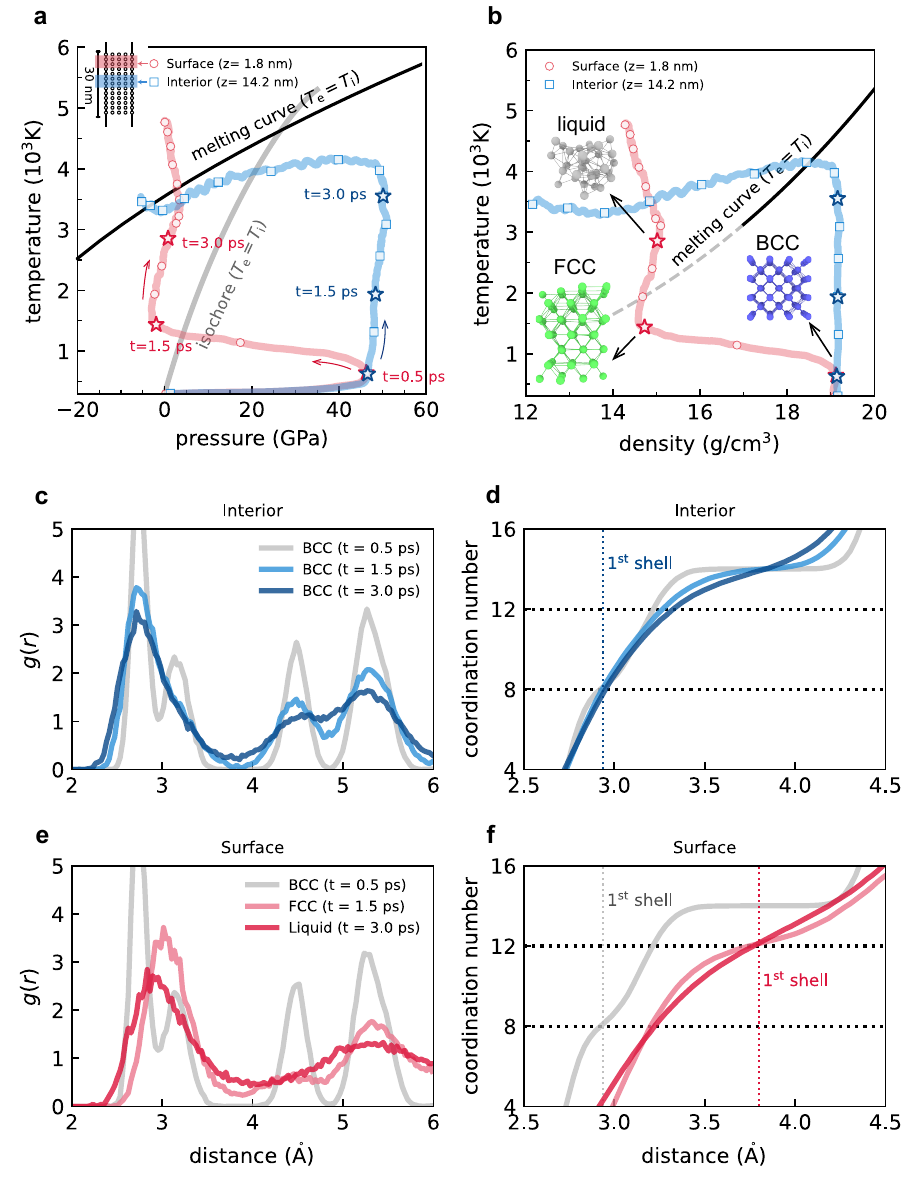} 
\caption{\textbf{Extreme heterogeneity in laser-excited W.} (a,b) Depth-dependent thermodynamic pathways in tungsten nanofilm under laser fluence of $120\ \rm{mJ\ cm^{-2}}$. Colored circles show thermodynamic states at 0.5 ps intervals, with arrows indicating evolution trajectories. Red stars mark three representative thermodynamic states of surface region ($z=1.8\ \rm{nm}$) corresponding to: initial electronic pressure buildup ($t=0.5\ \rm{ps}$), complete pressure release ($t=1.5\ \rm{ps}$), and surface melting ($t=3\ \rm{ps}$). Blue stars denote simultaneous thermodynamic states of interior region ($z=14.2\ \rm{nm}$) for comparison. The calculated equilibrium isochore (gray solid line) and melting curve (black solid line) are also presented for comparison \cite{zeng2023}. (c)\add{(e)} Radial distribution functions $g(r)$ and \add{(d)(f)} coordination numbers (CN) for surface and interior regions at selected times, where radial distribution of CN are obtained by integration of $g(r)$.}
\label{fig:2}
\end{figure}

The introduction of laser modification in PES through our ETD-NN model fundamentally alters this picture. As presented in Fig.\ref{fig:1}b, the W nanofilm quickly responds to laser heating on a sub-picosecond timescale, and exhibits instant surface expansion not observed in a purely thermal process. The heterogeneous nucleation of FCC phase is initially formed in the surface region, then the BCC-FCC transformation front moves inward with an average velocity of $2571\ \rm{m\ s^{-1}}$. Subsequently, as ion temperature increases above $\approx 2200\ \rm{K}$ at $t=2\ \rm{ps}$, although the temperature is significantly lower than the melting point ($T_{\rm m} = 3540\ \rm{K}$) under ambient condition, the collapse of crystal structure occurs in the surface region and proceeds inward as the BCC-FCC interface propagates. Such a melting process exhibits a well-defined solid-liquid interface moving inward, which is similar to the heterogeneous melting mechanism but has a high propagation speed up to $\approx 2500\ \rm{m\ s^{-1}}$. Here we name this melting behavior as “ultrafast heterogeneous melting” because the melting speed is an order of magnitude greater than that of conventional heterogeneous melting process ($\sim 10^2\ \rm{m\ s^{-1}}$) \cite{mo2018heterogeneous}. After $t = 4\ \rm{ps}$, the melt front rapidly advances, completing the melting process within the following two picoseconds.

\textbf{Electronic pressure-induced thermodynamic and structural heterogeneity.}
The ultrafast heterogeneous melting process exhibits spatial characteristics akin to conventional heterogeneous melting, with a well-defined melt front, while temporally resembling homogeneous melting due to its rapid progression. Such a unique combination of spatial and temporal features motivates a deeper exploration of the thermodynamic pathways and microscopic mechanisms involved. 

As shown in Fig.\ref{fig:2}, we present the thermodynamic evolution and microscopic structural transformation experienced by the interior ($z=14.2\ \rm{nm}$) and surface ($z=1.8\ \rm{nm}$) regions of the nanofilm. We observe that ultrafast electron heating induces extreme internal stress buildup within sub-picosecond timescales (Fig.\ref{fig:2}a). 
The hot electrons can contribute to an extra stress of $\approx 45\ \rm{GPa}$, which is significantly larger than that from thermal atomic vibrations of lattice. With the existence of free surface, a super-high expansion velocity of $\approx 755\ \rm{m\ s^{-1}}$ at the initial state ($t=0.4\ \rm{ps}$) is observed (see \add{Suppl. Fig.3}), and the uniaxial expansion process launches instantly to release this hot-electron-induced pressure. As shown in Fig.\ref{fig:2}b, the density in the surface region quickly decreases while the interior region is heated under isochoric condition during the first 3 ps. As a consequence, the inhomogeneous thermodynamic profiles are created. In density profile, a well-defined interface with sharp decrease of $\Delta \rho \approx 5\ \rm{g\ cm^{-3}}$ follows the release of stress (Fig.\ref{fig:2}b). This reduced density strongly influences the thermal stability of lattice and explains the early onset of surface disordering below equilibrium melting point. 

From a microscopic perspective, the relaxation of electronic stress waves induces local structural transformations. The uniaxial expansion increases interatomic distances perpendicular to the free surface, leading to a transformation from distorted BCC structure to FCC structure along the uniaxial Bain deformation path \cite{grimvall2012}. This observation is similar to the BCC-FCC transition observed by Murphy \textit{et al.} using isotropic isobaric \textit{ab initio} MD \cite{murphy2015}, highlighting the unique role of electronic pressure in modifying phase transitions. As shown in Fig.\ref{fig:2}\add{c-f}, the short-range order in the laser-excited nanofilm confirms the coexistence of high-density BCC and low-density FCC structures. At $t=1.5\ \rm{ps}$, the interior atoms display characteristics of a thermally-fluctuated BCC structure, while the surface atoms present typical FCC close-packed structures. This depth-variation in lattice coordination suggests that the density discontinuity driven by electronic pressure is accompanied by a disruption in local lattice symmetry. As the temperature continues to rise, the interplay between lattice heating and electronic pressure relaxation creates a multi-phase system composed of high-density BCC, low-density FCC, and disordered structures (\add{Suppl. Fig.4}).

\section{Discussion}

\begin{figure}[htbp]
\centering
\includegraphics[width=0.9\linewidth]{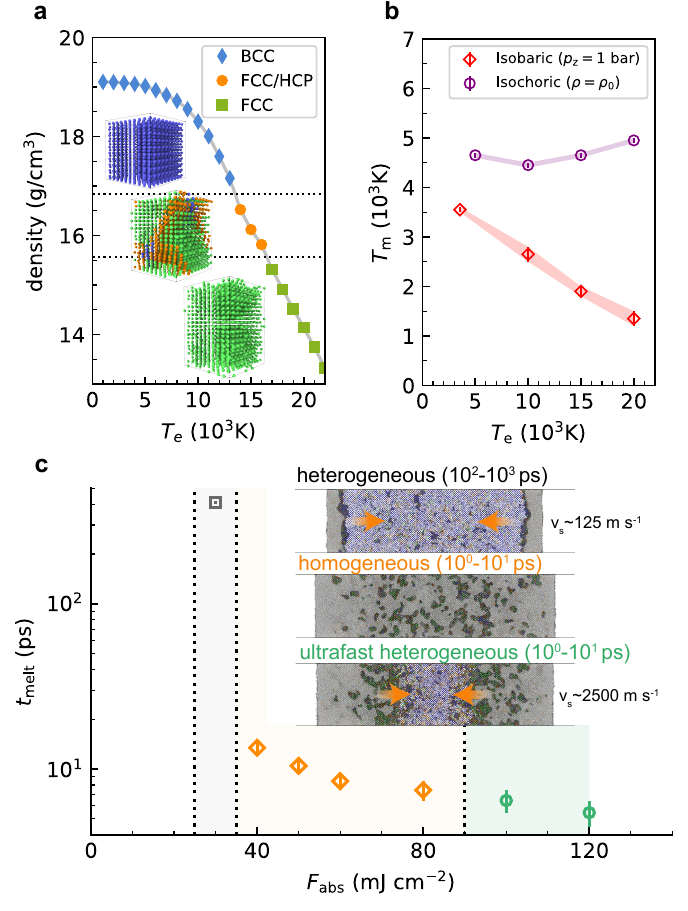}
\caption{\textbf{Laser fluence dependence of structural transformation in W. } (a) density decrease after electronic pressure relaxation along (100) direction, insets denote the uniaxially-distorted BCC, FCC with stacking faults, and FCC respectively. (b) isochoric and isobaric melting behavior under non-equilibrium condition, obtained from fixed-$T_{\rm e}$ DPMD simulations via two-phase method. \add{The error bar associated with the melting point is defined as half the temperature interval between the two-phase method simulations where the solid phase is stable and those where it is molten.} (c) complete melting time under different laser fluence. The yellow, red, and green region denotes the heterogeneous, homogeneous, and the ultrafast heterogeneous melting mechanism respectively. \add{The error bar for the complete melting time is defined as the temporal resolution of the atomic trajectory output from the TTM-DPMD simulations.}}
\label{fig:3}
\end{figure}

\textbf{Laser fluence dependence of structural transformation. }
To comprehensively understand the lattice response driven by electronic pressure,
we performed molecular dynamics simulations under uniaxial isobaric-isothermal ensemble ($T_{\rm i} = 300\ \rm{K}$, $p=1\ \rm{bar}$) across a wide range of $T_{\rm e}$ from $10,000\ {\rm K}$ to $20,000\ \rm{K}$. 

The equilibrated density and lattice structure is shown in Fig.\ref{fig:3}a. The structural transformation dynamics exhibits obvious dependence on laser fluence. 
Under relatively moderate two-temperature state ($T_{\rm e} \le 13,000\ \rm{K}$), the BCC structure maintains with slight uniaxial distortion. The density decreases from initial value of $19.15\ \rm{g\ cm^{-3}}$ to $17.15\ \rm{g\ cm^{-3}}$ at $T_{\rm e} = 13,000\ \rm{K}$, with a uniaxial strain of $\approx 8.81 \%$. 
As the $T_e$ increases, the electronic pressure drives the initial BCC structure towards the more close-packed structure, accompanied with shuffling of several close-packed atomic planes, thus obtaining FCC structure with stacking faults. When the $T_{\rm e}$ exceeds $16,000\ \rm{K}$, the stacking faults disappear and pure FCC structure is generated with a density of $14.13\ \rm{g\ cm^{-3}}$ at $T_{\rm e} = 20,000\ \rm{K}$, where the corresponding uniaxial strain is as high as $35.43 \%$. A series of TTM-DPMD simulations under different laser fluence confirmed the BCC-FCC phase transition threshold (\add{Suppl. Fig.5}), once the laser fluence exceeds $80\ \rm{mJ\ cm^{-2}}$, corresponding to a maximum electron temperature of $15,000\ \rm{K}$, the heterogeneous nucleation and growth of FCC structure begin to appear as the uniaxial expansion proceeds. 

Further, we investigate the laser fluence dependence of melting mechanism. By calculating the ultrafast electron diffraction pattern, we determine the complete melting time through the decay of (110) diffraction peak during laser heating processes (\add{Suppl. Fig.6}). For comparison, the results from purely thermal simulation are discussed (\add{Suppl. Fig.7}). 
As the laser fluence increases near the melting threshold ($53\ \rm{mJ\ cm^{-2}}$), the thermal heterogeneous melting starts from the free surface and proceeds slowly by the subsonic melt-front propagation ($\approx 125\ \rm{m\ s^{-1}}$), such process lasts hundreds of picoseconds. At high laser fluence, the homogeneous melting dominates, where the nucleation and growth of liquid region inside the foil quickly completes the melting within several picoseconds.

When incorporating the laser-excited PES through our ETD-NN model, the threshold fluences to the different melting regimes are reduced, both heterogeneous and homogeneous melting are predicted to occur more rapidly. 
Moreover, as shown in the Fig.\ref{fig:3}b, the release of electronic pressure introduces sharp decrease of local density, resulting in significantly reduced melting point in surface region as compared with isochoric condition. Therefore, as ion temperature increases, the low-density surface crystalline can quickly collapse into disordered structure before the interior melting.
Based on TTM-DPMD simulations (\add{Suppl. Fig.5}), here we can specify the laser fluence needed for triggering the newly discovered “ultrafast heterogeneous melting” process in the Fig.\ref{fig:3}c. As the initial electron temperature exceeds $T_{\rm e} = 16,000\ \rm{K}$, the isobaric melting point can deviate as large as $\Delta T_{\rm m} \approx 2750\ \rm{K}$ from isochoric one. Under such condition, the ultrafast heterogeneous process dominates the melting dynamics. 

\textbf{Nonthermal expansion signatures in X-ray diffraction lineouts. }
To elucidate the unique expansion dynamics under electronic pressure, we analyze the static structure factor $S(q)$, which directly correlates with X-ray diffraction (XRD) measurements. During the laser heating process, internal stress is released at the surface, generating stress waves that propagate inward. This results in a mixed signal representing both the expanded surface region and the isochoric interior region, as captured by $S(q)$.

\begin{figure}[htbp]
\centering
\includegraphics[width=1.0\linewidth]{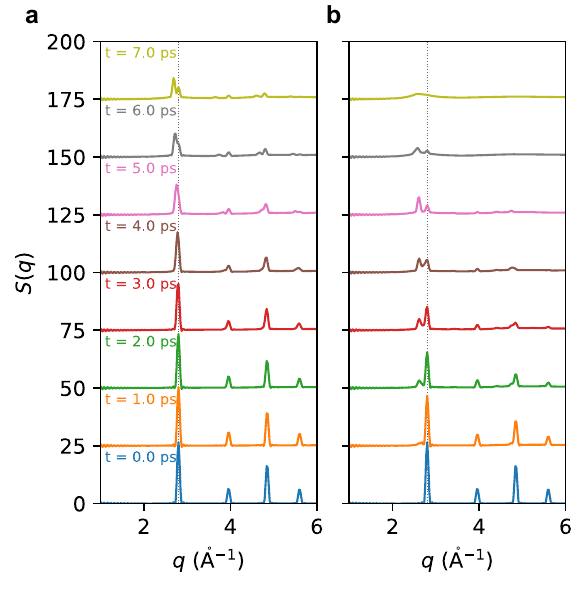}
\caption{Static structure factor $S(q)$ of laser-excited W nanofilm under laser fluence of $80\ \rm{mJ\ cm^{-2}}$, estimated from TTM-DPMD trajectories with (a) ground-state PES (b) laser-excited PES. The black line highlights the first diffraction peak from initial BCC structure.}
\label{fig:xrd}
\end{figure}

\begin{figure}[htbp]
\centering
\includegraphics[width=1.0\linewidth]{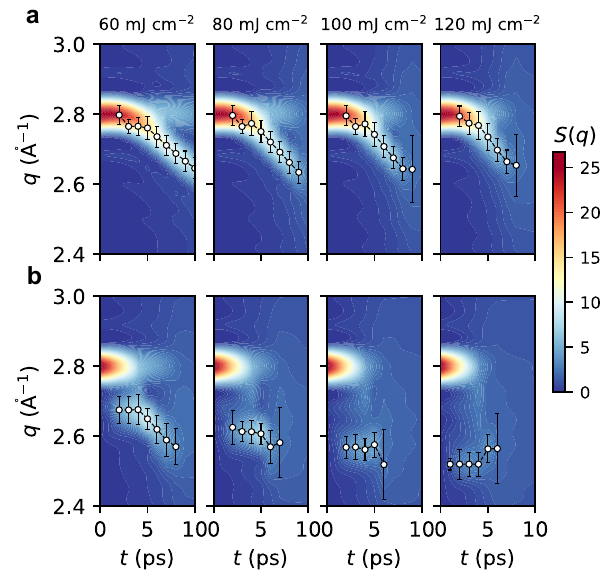}
\caption{Temporal evolution of static structure factor $S(q)$ of laser-excited W nanofilm at different laser fluence, estimated from TTM-DPMD trajectories with (a) ground-state PES (b) laser-excited PES. The positions of the new diffraction peak, \add{estimated by Gaussian fitting}, are marked by black circles to highlight the difference between thermal and nonthermal expansion dynamics. \add{The error bars represent the standard deviation of the Gaussian peak width.}
}
\label{fig:xrd-all}
\end{figure}

As shown in Fig.\ref{fig:xrd}, we take laser-excited W with laser fluence of $F_{abs}=80\ \rm{mJ\ cm^{-2}}$ as an example. In the purely thermal process, the dominant role of electron-phonon coupling leads to the accumulation of thermal kinetic pressure inside the foil, thus driving gradual surface expansion. 
From Fig.\ref{fig:xrd}a, the new characteristic peaks splits from the original peak at $q= 2.8\ \rm{\mathring A^{-1}}$, then continuously shifts to lower $q$ values, corresponding to the gradual increase in interatomic distance of surface region under uniaxial expansion.

In nonthermal processes (Fig.\ref{fig:xrd}b), the generation of electronic pressure originates from the formation of the initial two-temperature state. Consequently, the release of hot-electron-induced pressure occurs at sub-picosecond timescales, resulting in significant uniaxial distortion. At $t=2\ \rm{ps}$, the proportion of low-density surface region reaches a detectable level for XRD, resulting in a notable splitting of the first diffraction peak.
A new characteristic peak has been observed at $q_{\rm s}=2.613\ \rm{\mathring A^{-1}}$.
Unlike thermal processes, the position of this new peak $q_{\rm s}$ remains unchanged over time. Only its intensity gradually increases, indicating the ongoing nonthermal expansion process. 
As the lattice temperature increases, the long-range structure of the lattice progressively dissipates, leading to the disappearance of characteristic peaks associated with crystalline structure in $S(q)$ at $t=7\ \rm{ps}$.

As shown in Fig.\ref{fig:xrd-all}, we present the temporal evolution of $S(q)$ under various laser fluence, ranging from $60\ \rm{mJ\ cm^{-2}}$ to $120\ \rm{mJ\ cm^{-2}}$, aiming to provide a comprehensive understanding of expansion dynamics. Analysis of the $2.4-3.0\ \rm{\mathring A^{-1}}$ diffraction regime reveals distinct fluence-dependent behaviors: (1) During thermal expansion, it manifests as continuous shifting, where a higher lattice heating rate can result in a slightly faster shift of the new diffraction peak, as shown in Fig.\ref{fig:xrd-all}a. (2) During nonthermal expansion process, it exhibits as discontinuous splitting, and the position of the new diffraction peak displays a strong dependence on laser fluence. A larger electronic pressure corresponds to a lower $q$ value position of the new peak, as illustrated in Fig.\ref{fig:xrd-all}b. The observed qualitative disparity in expansion dynamics, as indicated by $S(q)$, implies the possibility of identifying evidence of electronic pressure relaxation dynamics through time-resolved X-ray diffraction experimental methodologies.

\textbf{Influence of electron-phonon coupling strength and sample thickness.} 
We further examined the effects of electron-phonon coupling strength $G(T_{\rm e})$ and sample thickness on the ultrafast heterogeneous melting behavior. Although theoretical predictions for $G(T_{\rm e})$ at high electron temperatures vary by orders of magnitude \cite{lin2008, medvedev2020electron, smirnov2025}, and the sample thickness modulates the competition between electronic pressure relaxation and thermalization, our simulations show that these factors only quantitatively shift the melting thresholds and does not change the heterogeneous melting behaviors (\add{Suppl. Fig.8-10}). Specifically, adopting a stronger electron-phonon coupling model ($G(T_{\rm e})$ from Lin \textit{et al.} or Smirnov \textit{et al.}) leads to a lower laser fluence threshold required to trigger ultrafast heterogeneous melting, as the lattice heats up more rapidly. While using a weaker coupling model (e.g., $G(T_{\rm e})$ from Medvedev \textit{et al.}) yields melting dynamics and thresholds similar to those obtained with a constant electron-phonon coupling.

Similarly, reducing the sample thickness shortens the propagation time of electronic stress waves, and still results in the emergence of pressure-driven heterogeneous melting fronts even in films as thin as $10\ \rm{nm}$. For intermediate thickness such as $20\ \rm{nm}$, our simulations show that the propagation time of stress waves becomes comparable to the thermalization timescale, leading to a competition between ultrafast heterogeneous melting and homogeneous melting. These results highlight that the emergence of electronic pressure from the laser-excited PES fundamentally modifies the melting dynamics, and this effect manifests similarly across a wide range of model parameters and sample geometry.

\textbf{Ubiquity of heterogeneous structural transformations. }
To demonstrate the ubiquity of electronic pressure in laser-driven processes, we extend the TTM-DPMD framework to gold (Au), a noble metal with fully occupied $d$-orbitals. Despite its distinct electronic structure, Au exhibits analogous nonthermal dynamics (\add{Suppl. Fig.11}).
At elevated electron temperatures, the hot electrons also contribute non-negligible electronic pressure in Au, reaching 25 GPa at $T_{\rm e}=20,000\ \rm{K}$. 
Under isobaric-isothermal condition ($T_{\rm i}=300\ \rm{K}$, $p_{\rm z}=1\ \rm{bar}$), the laser-excited Au undergoes significant uniaxial distortion. As $T_{\rm e}$ reaches 17,000 K, the expansion can even destabilize the lattice, triggering a disordered transition at room temperature ($T_{\rm i}=300\ \rm{K}$)—a “nonthermal melting” regime where amorphization occurs without lattice heating. These findings align with the conclusions of Daraszewicz \textit{et al.}, and Medvedev \textit{et al.} regarding nonthermal behaviors under isobaric constraints \cite{daraszewicz2013, medvedev2020}, suggesting that the ultrafast disordering of laser-excited Au is primarily due to electron pressure relaxation process.

Further, large-scale simulations of 35-nm-thick free-standing Au nanofilm confirm ultrafast heterogeneous melting (\add{Suppl. Fig.12}), mirroring W’s behavior. 
With the release of electronic pressure, the surface atoms directly transforms into a disordered structure due to significant reduction in melting point. The melt front then moves inward at a speed of $2,916\ \rm{m\ s^{-1}}$, completing the melting process within $8\ \rm{ps}$. These results, along with previous prediction on the widespread existence of electronic pressure in laser-excited materials \cite{bevillon2014}, underscore the pervasive nature of electron pressure relaxation dynamics and ultrafast heterogeneous structural transformation in laser-driven processes.

\textbf{Summary.} By investigating real-time response of laser-excited W and Au nanofilm via TTM-DPMD approach, we unraveled the vital role of the electronic pressure in the laser-driven dynamics, which was previously not fully understood. As an integral part of nonthermal behavior, the electronic pressure builds up simultaneously with appearance of the two-temperature state.
At elevated electron temperatures, the additional pressure contributed by hot electrons becomes significant, reaching up to $p_{\rm e} \sim 10^1\ \rm{GPa}$ at $T_{\rm e}\sim 10^4\ \rm{K}$. 
The release of electronic pressure triggers violent uniaxial expansion from free surface, which exhibits a drastic difference from thermal expansion in temporal behaviors, which can be identified by time-resolved X-ray diffraction techniques. As a result, pressure, density, and temperature discontinuity in thermodynamic profile is generated and moves inward at a high speed on the order of $10^3\ \rm{m\ s^{-1}}$.

The following structural transformation is therefore dominated by the relaxation of electronic pressure. Two unexpected phenomena, absent in either purely thermal process or previous lattice dynamics calculations, were observed: (i) heterogeneous solid-solid transition (in W), which directly arises from the expansion process that drives the BCC-FCC transition along the uniaxial Bain path. (ii) ultrafast heterogeneous melting (both in W and Au), which is a combination of density decrease driven by nonthermal expansion and temperature increase induced by electron-phonon coupling. These findings suggest a extreme structural heterogeneity in laser-excited metal, which has a profound influence on the interpretation of experiments and our comprehension of nonthermal behavior.

We also note that this structural transformation is fundamentally different from the conventional nonthermal phase transition driven by lattice instability. The latter is expected to occur in less than a phonon period \cite{zijlstra2008, wall2018, wang2020}, which can be considered as homogeneous and ultrafast. Instead, the nonthermal expansion induces heterogeneous structural transformation, which can strongly rely on the crystal symmetry, the orientation of free surface, and the strength of nonthermal pressure. Especially, even when phonon spectra suggest a dynamically stable structure, this structural transformation can occur. This is particularly evident in laser-excited Au.

Moreover, considering the interplay between this heterogeneous phase transition and ultrafast melting, whether the intermediate solid phase can be observed by experimental measurements depends on accurate determination of electron-phonon coupling. Especially for the extremely inhomogeneous thermodynamic profile, both electron temperature and density dependence of electron-phonon energy exchange rate should be considered in the future work. On this issue, TTM-DPMD approach can to be coupled with the newly-developed deep learning scheme to infer electronic structure on-the-fly \cite{zeng2022towards, li2022deep}.

\section{Methods}
\textbf{Electron temperature dependent neural network model. }
In laser-excited metals, electron-electron scattering occurs on the femtosecond timescale \cite{chen2021ultrafast}, leading to ultrafast thermalization of excited electrons within $10^1-10^2\ \rm{fs}$. This results in a thermal equilibrium state of electron subsystem with a well-defined temperature $T_e$. Since we focus on the ion dynamics for $\rm{ps}$ timescale of interest, the nonthermal nature of laser-excited electrons in the early stages after excitation is not considered here. In this context, 'nonthermal' primarily denotes the non-equilibrium state where the electrons have a well-defined temperature $T_{\rm e}$ that is different from the lattice temperature $T_{\rm i}$. Therefore, the complex potential energy surface (PES) of laser-excited matter can be described within the framework of finite-temperature density functional theory \cite{mermin1965}, which is a free energy surface $A(\mathbf R, T_{\rm e})$ of an electron-ion system.

To capture the modification on PES introduced by hot electrons, here we adopt the deep neural network with an additional parameter, electron temperature $T_e$, named as electron-temperature-dependent deep potential model \cite{zeng2023}, 
\begin{align}
A = A(\mathbf R, T_e) = \sum_i \mathbf N_{\alpha_i} (\mathbf D_{\alpha_i}(r_i,\{r_j\}_{j\in n(i)}), T_e)
\end{align}
where $A(\mathbf R, T_e)$ indicates that the free energy depends on the local atomic environment $\mathbf R$ and electron temperature, $\mathbf N_{\alpha_i}$ denotes the neural network of specified chemical species of $\alpha_i$ of atom $i$, and the descriptors $\mathbf D_{\alpha_i}$ describes the local environment of atom $i$ with its neighbor list $n(i)=\{j|r_{ji}<r_{\rm{cut}}\}$. 
The ETD-NN model can inherently incorporate the nonthermal characteristics of laser-excited matters in molecular dynamics simulations, including the hot-electron-modulated atomic forces and virial tensor, 
\begin{align}
\mathbf F &= -\frac{\partial A}{\partial \mathbf R_i} = -\frac{\partial U}{\partial \mathbf R_i} +T_e \frac{\partial S}{\partial \mathbf R_i} \\
\mathbf \Xi &= \frac{\partial A}{\partial \mathbf \Omega} \cdot \mathbf \Omega^T 
= \frac{\partial U}{\partial \mathbf \Omega} \cdot \mathbf \Omega^T - T_e \frac{\partial S}{\partial \mathbf \Omega} \cdot \mathbf \Omega^T
\end{align}
here $\mathbf R_i$ denotes the coordinate vector of atom $i$, $\mathbf \Omega$ denotes a $3\times 3$ matrix formed by three lattice vectors and $\mathbf \Omega^T$ is its transpose. Consider a system composed of $N$ atoms, the total pressure of this electron-ion system is determined by the summation of ionic thermal kinetic contribution and virial contribution.

In the maintext, the ETD-NN model of tungsten is generated with DeePMD-kit packages \cite{dpkitv2} and DP-Generator \cite{zhang2019active}. The configurations space with different electronic occupation considered in this work is efficiently sampled.
Based on the training data set generated in previous work \cite{zeng2023}, \add{specifically, BCC structure (54 atoms) and liquid structure (54 atoms) are considered as the initial configurations and run DPMD under NVT and NPT ensemble (both isotropic and uniaxial constrains are considered), where temperatures ranges from 100 K to 6000 K, pressure ranges from -15 to 60 GPa, and corresponding electronic temperature ranges from 100 K to 25000 K. The training sets consist of 6366 configurations under equilibrium condition ($T_e = T_i$) and 6820 configurations sampled under two-temperature state ($T_e > T_i$).} 
Moreover, we further improved the uniaxial expansion behavior of laser-excited tungsten along (100), (110) and (111) direction by sampling additional 1,860 configurations, where the ion temperature ranges from 100 K to 4,000 K, electron temperature ranges from 100 K to 25,000 K. 

\add{For DP training, the embedding network is composed of three layers (25, 50, and 100 nodes) while the fitting network has three hidden layers with 240 nodes in each layer. The total number of training steps is set to 400 000. The radius cutoff $r_c$ is chosen to be $6.0\ {\rm \mathring A}$. The weight parameters in loss function for energies $p_e$, forces $p_f$, and virials $p_V$ are set to $(0.02, 1000, 0.02)$ at the beginning of training and gradually change to $(1.0, 1.0, 1.0)$.}

The self-consistency calculations are all performed with the VASP packages \cite{kresse1996efficient}. The Perdew-Bruke-Erzerhof (PBE) exchange correlation functional is used \cite{perdew1996generalized} and psudopotential takes the projector augmented-wave (PAW) formalism \cite{blochl1994projector}. The sampling of Brillouin zone is chosen as 0.2 $\rm{\mathring A^{-1}}$ under ambient condition and 0.5 $\rm{\mathring A^{-1}}$ for high temperature ($T\ge 1600\ \rm{K}$).

\textbf{Two-temperature model coupled neural network molecular dynamics approach.}
To simulate the real-time response of material upon laser excitation, the additional description on electron subsystem is introduced and strongly coupled with atomic system in TTM-DPMD approach.
The heat conduction equation of electron continuum characterizes the temporal evolution of electron occupation, thus governing the transition of atomic system between different $T_e$-dependent PES. Langevin dynamics is incorporated to mimic the dynamic electron-ion collision, thus including the non-adiabatic energy exchange between electron and atomic subsystem,
\begin{align}
C_e(T_e)\frac{\partial T_e}{\partial t} &= \nabla \cdot(\kappa_e \nabla T_e) - g_{ei}(T_e) (T_e-T_i) + S(\mathbf r, t )\\
m_i\frac{d^2 \mathbf r_i}{dt^2} &=-\nabla_i A(T_e) -\gamma_i \mathbf v_i + \tilde {\mathbf F_i}(t)
\end{align}
where $C_e$ the electron heat capacity, $\kappa_e$ the electronic thermal conductivity, $g_{ei}$ the electron-phonon coupling constant, $S(\mathbf r,t)$ the laser source, $\gamma_i$ the friction parameter, $\tilde {\mathbf F_i}$ the random force with Gaussian distribution.

To simulate the material response that relevant to real experimental conditions, 
here the 30-nm-thick W nanofilm \cite{mo2019} and 35-nm-thick Au nanofilm \cite{mo2018heterogeneous} are chosen as the target sample. A Gaussian temporal profile of laser pulse is used, and the duration is set to $130\ \rm{fs}$.
Since both the thickness of W and Au nanofilm is comparable to the mean free path of excited electrons ($\sim 33\ \rm{nm}$ in W and $\sim 100\ \rm{nm}$ in Au), the ballistic motion of the excited electrons leads to the fast (within $\sim 100\ \rm{fs}$) redistribution of the deposited energy within the ballistic range. As a result, the uniform laser energy deposition is expected, 
\begin{align}
S(t)=\frac{F_{\rm abs} }{\sigma d \sqrt{2 \pi}} \exp(-\frac{(t-t_0)^2}{2\sigma^2})
\end{align}
where $F_{\rm abs}$ the absorbed laser fluence, $\sigma$ the standard deviation of the temporal Gaussian distribution, $d$ the thickness of nanofilm, $t_0$ the time zero defined as the arrival of the maximum of the laser pulse. 
The absorbed laser energy density $\epsilon$ can be estimated via $\epsilon = F_{abs} / \rho d$, where $\rho$ the mass density.

In the maintext, the TTM-DPMD simulations is performed with LAMMPS packages \cite{plimpton1995fast} through modified EXTRA-FIX module. The electronic heat capacity is obtained by individual DFT calculations $C_e=T_e\frac{\partial S_e}{\partial T_e}$. The electron thermal conductivity is described by the Drude model relationship, $\kappa_e(T_e,T_i)=\frac{1}{3} v_F^2 C_e(T_e) \tau_e(T_e,T_i)$, where $v_F$ is Fermi velocity and $\tau_e(T_e,T_i)$ is the total electron scattering time defined by the electron-electron and electron-phonon scattering rates, $1/\tau_e=1/\tau_{e-e}+1/\tau_{e-ph}=AT_e^2 +BT_i$. The coefficients $A=2.11\times10^{-4}\ \rm{K^{-2}\ ps^{-1}}, B=8.4\times 10^{-2}\ \rm{K^{-1}\ ps^{-1}},v_F=9710\ \rm{\mathring A\ ps^{-1}}$ are adopted.
Considering the accurate determination of electron-phonon coupling at elevated electron temperatures remains controversial, a constant electron-phonon coupling value ($G_0=2.0\times10^{17}\ \rm{W\ m^{-3}\ K^{-1}}$) from Mo \textit{et al.}'s experiments \cite{mo2019} is used in the main text. Furthermore, a  detailed discussion with the different electron-temperature dependent $G(T_e)$ predictions \cite{lin2008, medvedev2020electron, smirnov2025} is also provided, we refer to the Discussion section and Supplementary Materials (\add{Suppl. Fig.8-9}).

For atomic system, the sample geometry of 30-nm-thick 100-oriented single-crystalline W foil is a parallelepiped $L_x\times L_y \times L_z$ with $L_x = L_y = 30 a_0$ and $L_z=95a_0$ (171,000 atoms). $a_0 = 3.17104\ \rm{\mathring A}$ is the parameter of the elementary BCC cell for W corresponding to the ETD-NN model prediction at 300 K and 0 GPa. Periodic boundary conditions are applied in the x and y directions, while free boundary condition is applied to $z$-axis to allow the expansion of free surface. The lateral periodic conditions ($x$ and $y$-axis) simulate the experimental situation in which the laser spot diameter is large (hundreds of microns) compared to the depth of the laser energy deposition (tens of nanometers) so that the effects of the edges of the laser beam can be neglected.

\section{Data availability}
\add{The relevant data generated in this study have been deposited in the Zenodo \cite{SourceData}.}

\section{Code availability}
\add{LAMMPS and DeePMD-kit are free and open source codes available at https://github.com/lammps/lammps and https://github.com/deepmodeling/deepmd-kit, respectively. VASP is a commercial code available from https://www.vasp.at. Detailed instructions for obtaining and using these codes can be found on their respective websites.}

\bibliography{reference}

\section{Acknowledgments}
This work was supported by the National Natural Science Foundation of China under Grant Nos. 12304307, 12122103, 12104507, \add{12504326, 12534013}, the Science and Technology Innovation Program of Hunan Province under Grant No. \add{2025YJ001} and 2021RC4026.

\section{Author Contributions Statement}
Q.Z and J. D. proposed the original idea and designed the research. Q. Z. carried out the simulations, Q. Z., X. Y., D. K., and J.D. analyzed and interpreted the results. Q. Z, X. Y. and J. D. wrote the manuscript with the help from other authors.
B. C., S. Z., and K. C. provided additional support for the interpretation of the results.

\section{Competing Interests Statement}
The authors declare no competing interests.


\newpage

\setcounter{figure}{0}
\begin{figure*}[htbp]
\centering
\caption{\textbf{Diverging melting dynamics in laser-excited tungsten.} Under absorbed laser fluence of $120\ \rm{mJ\ cm^{-2}}$,  (a) Conventional TTM-MD prediction with ground-state PES $A(\mathcal R)$ showing homogeneous melting \add{governed} by electron-phonon coupling. (b) Our TTM-DPMD results with laser-excited PES $A
(\mathcal R, T_{\rm e})$ revealing electronic pressure-driven heterogeneous melting. The local structures are identified by the polyhedral template 
matching (PTM) method \cite{larsen2016}, and polyhedral surface meshes around FCC-type (green) 
and amorphous-type (gray) particles are constructed to highlight the heterogeneity in lattice symmetry. Red arrows indicate the propagation direction of phase transition (solid-liquid or BCC-FCC) interfaces. \add{The atomic configurations are visualized by OVITO software \cite{ovito}.}} 
\label{fig:1}
\end{figure*}

\begin{figure}[htbp]
\centering
\caption{\textbf{Extreme heterogeneity in laser-excited W.} (a,b) Depth-dependent thermodynamic pathways in tungsten nanofilm under laser fluence of $120\ \rm{mJ\ cm^{-2}}$. Colored circles show thermodynamic states at 0.5 ps intervals, with arrows indicating evolution trajectories. Red stars mark three representative thermodynamic states of surface region ($z=1.8\ \rm{nm}$) corresponding to: initial electronic pressure buildup ($t=0.5\ \rm{ps}$), complete pressure release ($t=1.5\ \rm{ps}$), and surface melting ($t=3\ \rm{ps}$). Blue stars denote simultaneous thermodynamic states of interior region ($z=14.2\ \rm{nm}$) for comparison. The calculated equilibrium isochore (gray solid line) and melting curve (black solid line) are also presented for comparison \cite{zeng2023}. (c)\add{(e)} Radial distribution functions $g(r)$ and \add{(d)(f)} coordination numbers (CN) for surface and interior regions at selected times, where radial distribution of CN are obtained by integration of $g(r)$.}
\label{fig:2}
\end{figure}

\begin{figure}[htbp]
\centering
\caption{\textbf{Laser fluence dependence of structural transformation in W. } (a) density decrease after electronic pressure relaxation along (100) direction, insets denote the uniaxially-distorted BCC, FCC with stacking faults, and FCC respectively. (b) isochoric and isobaric melting behavior under non-equilibrium condition, obtained from fixed-$T_{\rm e}$ DPMD simulations via two-phase method. \add{The error bar associated with the melting point is defined as half the temperature interval between the two-phase method simulations where the solid phase is stable and those where it is molten.} (c) complete melting time under different laser fluence. The yellow, red, and green region denotes the heterogeneous, homogeneous, and the ultrafast heterogeneous melting mechanism respectively. \add{The error bar for the complete melting time is defined as the temporal resolution of the atomic trajectory output from the TTM-DPMD simulations.}}
\label{fig:3}
\end{figure}

\begin{figure}[htbp]
\centering
\caption{Static structure factor $S(q)$ of laser-excited W nanofilm under laser fluence of $80\ \rm{mJ\ cm^{-2}}$, estimated from TTM-DPMD trajectories with (a) ground-state PES (b) laser-excited PES. The black line highlights the first diffraction peak from initial BCC structure.}
\label{fig:xrd}
\end{figure}

\begin{figure}[htbp]
\centering
\caption{Temporal evolution of static structure factor $S(q)$ of laser-excited W nanofilm at different laser fluence, estimated from TTM-DPMD trajectories with (a) ground-state PES (b) laser-excited PES. The positions of the new diffraction peak, \add{estimated by Gaussian fitting}, are marked by black circles to highlight the difference between thermal and nonthermal expansion dynamics. \add{The error bars represent the standard deviation of the Gaussian peak width.}
}
\label{fig:xrd-all}
\end{figure}

\end{document}